\documentclass[prl,showpacs,twocolumn,superscriptaddress,floatfix,10pt]{revtex4-2}
\usepackage{setspace} 
\usepackage[utf8]{inputenc}
\usepackage{float}
\usepackage{amsmath,amssymb,amsfonts,bm}
\usepackage{graphicx}
\usepackage{multirow}
\usepackage[usenames,dvipsnames]{color}
\allowdisplaybreaks
\usepackage[
colorlinks=true,
linkcolor=blue,
breaklinks=true,
urlcolor=blue,
citecolor=blue]{hyperref}
\usepackage{orcidlink}

\newcommand{\be}{\begin{equation}} 
\newcommand{\ee}{\end{equation}}
\newcommand{\bea}{\begin{eqnarray}} 
\newcommand{\eea}{\end{eqnarray}}

\newcommand{\nno}{\nonumber\\}

\newcommand{\im}{\operatorname{Im}}
\newcommand{\re}{\operatorname{Re}}

\definecolor{green}{rgb}{0,0.6,0}

\newcommand{\dd}{\text{d}}

\newcommand{\tg}{\tilde{g}}

\newcommand{\kk}{k^{\prime 2}}
\newcommand{\mex}{m_\text{ex}^2}

\newcommand{\deta}{\eta}

\newcommand{\uestc}{\affiliation{School of Physics, University of Electronic Science and Technology of China, Chengdu 611731, China}}

\newcommand{\itp}{\affiliation{Institute of Theoretical Physics, Chinese Academy of Sciences, Beijing 100190, China}}

\newcommand{\ucas}{\affiliation{School of Physical Sciences, University of Chinese Academy of Sciences, Beijing 100049, China}}

\newcommand{\peng}{\affiliation{Peng Huanwu Collaborative Center for Research and Education, Beihang University, Beijing 100191, China}}

\begin{document}
\title{Effective range expansion with a long-range force}

\begin{abstract}

The validity range of the time-honored effective range expansion can be very limited due to the presence of a left-hand cut close to the two-particle threshold. Such a left-hand cut arises in the two-particle interaction involving a light particle exchange with a mass small or slightly heavier than the mass difference of the two particles, identified as a long-range force, a scenario encountered in a broad range of systems. This can hinder a precise extraction of low-energy scattering observables and resonance poles. To address this issue, we propose a new parameterization for the low-energy scattering amplitude that accounts for the left-hand cut. The parameterization is like a Pad\'e approximation but with nonanalytic terms from the left-hand cut and can be regarded as an extension of the effective range expansion. This parameterization is versatile and applicable to a broad range of systems with Yukawa-type interactions---including particle, hadronic, nuclear, cold atom and quantum gas systems.
In particular, it should be invaluable in understanding various near-threshold hadron resonances. As byproducts, we also show that the parameterization can be used to extract the couplings of the exchanged particle to the scattering particles, and derive expressions for amplitude zeros caused by the interplay between the short- and long-range interactions. 

\end{abstract}

\author{Meng-Lin Du\orcidlink{0000-0002-7504-3107}}\email{du.ml@uestc.edu.cn}
\uestc

\author{Feng-Kun~Guo\orcidlink{0000-0002-2919-2064}}\email{fkguo@itp.ac.cn}
\itp \ucas \peng

\author{Bing Wu\orcidlink{0009-0004-8178-3015}}\email{wu.bing@uestc.edu.cn}
\uestc

\maketitle

{\it Introduction.}---Effective range expansion (ERE)~\cite{Bethe:1949yr,Blatt:1949zz} is a powerful tool extensively utilized in, particularly, hadronic, nuclear, and cold atom physics to describe the behavior of scattering amplitudes at low energies. It provides an efficient parameterization to extract dynamical information regarding the interaction between two particles in the low-energy regime. ERE is based on the idea that, at low energies, the short-range interactions can be effectively described by a series expansion in terms of the relative momentum of the scattering particles. It is a direct consequence of unitarity and analyticity of the amplitude in the vicinity of the two-particle threshold. Unitarity dictates that the imaginary part of the inverse of a  two-particle elastic scattering amplitude $f$ is given by 
\begin{align}
\im\frac{1}{f(E)} = -\frac{2\pi}{\mu }\im \frac{1}{T(E)} = -k,
\label{eq:invf}
\end{align}
in the nonrelativistic kinematics. Here $T$ denotes the $T$-matrix, $k=\sqrt{2\mu E}$ with $\mu$ the reduced mass of the scattering particles and $E$ the center-of-mass (c.m.) energy of the system relative to the two-particle threshold. Analyticity of the amplitude implies that in the vicinity of the two-particle threshold, the real part of the inverse amplitude can be expanded in powers of $k^2=2\mu E$~\cite{Bethe:1949yr,Blatt:1949zz}. For the $S$ wave, ERE reads
\begin{align}
    \label{eq:ere0}
\re\frac{1}{f(E)} = k\cot \delta = \frac{1}{a}+\frac12 r k^2 + \mathcal{O}(k^4),
\end{align}
where $\delta$ is the scattering phase shift, and the parameters $a$ and $r$ are the scattering length and effective range, respectively. However, the convergence radius of ERE~\eqref{eq:ere0} is bounded by the location of the nearest singularity in the energy plane
(for an extension of ERE involving unstable particles, see Refs.~\cite{Braaten:2009jke, Baru:2021ldu}). Among these singularities, the left-hand cut (lhc) generated by the Yukawa-type exchange of a light particle with a mass small or slightly heavier than the mass difference of the two scattering particles is particularly significant as its lhc branch point lies in the vicinity of the threshold, limiting the applicability of Eq.~\eqref{eq:ere0} to a very small range. Such an exchange corresponds to a force that operates over a long but still finite range.

In hadron physics, in the past two decades, tens of resonances have been observed experimentally in the charmonium and bottomonium mass ranges (for recent reviews, see, e.g., Refs.~\cite{Hosaka:2016pey, Esposito:2016noz, Guo:2017jvc, Olsen:2017bmm, Brambilla:2019esw, Chen:2022asf, Meng:2022ozq, ParticleDataGroup:2024cfk}). Many of them are close to two-hadron thresholds, making the corresponding amplitudes potentially suitable for exploration using ERE. However, the applicability of ERE is hindered mainly by two obstacles. One is the coupled-channel effects, and the other is the proximity of threshold to the lhc resulting from the one-particle exchange (usually the one-pion exchange), which occurs in a wide range of scatterings such as baryon-(anti)baryon, $B^{(*)}B^*$, $B^{(*)}\bar{B}^*$, $ND^{*}$, $N\bar D^{*}$, $\Sigma_{(c)}^{(*)}D^{*}$, $\Sigma_{(c)}^{(*)}\bar D^{*}$, and so on. 
The first obstacle may be addressed by constructing coupled-channel nonrelativistic effective field theories as in, e.g., Refs.~\cite{Cohen:2004kf,Braaten:2007nq,Lensky:2011he, Dong:2020hxe, Sone:2024nfj, Zhang:2024qkg}. The second obstacle, the lhc, is the focus of this Letter. 

Experimental observations also stimulated the lattice chromodynamics (QCD) community to investigate the interactions of hadrons near the thresholds. 
For instance, the $DD^*$ and $D\bar{D}^*$ systems have been extensively investigated in lattice QCD~\cite{Prelovsek:2013cra,Prelovsek:2014swa,HALQCD:2016ofq,Cheung:2017tnt,CLQCD:2019npr,Piemonte:2019cbi,Padmanath:2022cvl,Lyu:2023xro,Li:2024pfg,Sadl:2024dbd}.
In these calculations, one common strategy is to use the finite-volume L\"uscher formalism to extract the ERE parameters~\cite{Prelovsek:2013cra,Prelovsek:2014swa,Piemonte:2019cbi,Padmanath:2022cvl,Li:2024pfg,Sadl:2024dbd}, from which the pole positions are then extracted.
However, with the quark masses used in the lattice calculations, there can be lhcs near the thresholds. 
It is crucial to highlight that the presence of the lhc can induce significant physical effects such as severe modification of the phase shift and pole structure, as discussed in Ref.~\cite{Du:2023hlu} (see also Ref.~\cite{Dawid:2023jrj}). 
Various finite volume formalisms were proposed to remedy the issue~\cite{Meng:2023bmz, Raposo:2023oru,Hansen:2024ffk,Bubna:2024izx}.
In this Letter, we propose a new ERE parameterization for the low-energy $S$-wave amplitude that accounts for the lhc.
The parameterization can be regarded as an extension of ERE, and we will show that with only a few parameters, it can accurately describe the low-energy scattering amplitude. 

\begin{figure}[tb]
    \begin{center}
    \includegraphics[width=0.9\linewidth]{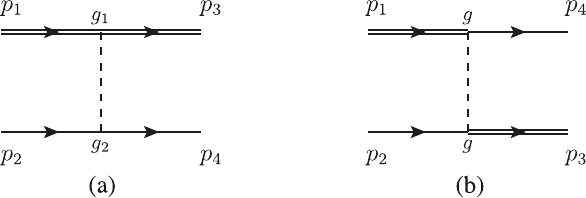}
    \caption{Two-particle scattering through (a) $t$- and (b) $u$-channel one-particle exchanges. Here, $g_{1,2}$ and $g$ denote the coupling constants, and $p_i$'s are the momenta of particles. }
    \label{fig:feyndiag}
    \end{center}
    \end{figure}

{\it The left-hand cut.}---The lhc is a general feature of the partial wave amplitude stemming from the crossed-channel singularity, and the branch point of the lhc is set by that the singularity occurs at the endpoint of the partial wave projection. A significant impact on observables can arise from the lhc if the two particles can interact by exchanging a nearly on-shell particle, see Fig.~\ref{fig:feyndiag}. 
We focus only on the lhc generated by this type of one-particle exchange (OPE) potential, corresponding to a long-range interaction, as it is the nearest lhc at low energies. The discontinuity of the amplitude across the lhc is induced by the logarithm in the partial wave of the OPE. For an $S$-wave elastic scattering, the lhc parts of the $t$- and $u$-channel exchange partial waves are given by
\begin{align}\label{eq:L}
L_t(k^2) &\equiv\frac12\int_{-1}^{+1}\frac{\dd\cos\theta}{t-\mex} \,= -\frac{1}{4k^2}\log\frac{\mex/4+k^2}{\mex/4}, \nno
L_u(k^2) &\equiv \frac12\int_{-1}^{+1}\frac{\dd\cos\theta}{u-\mex} 
\approx-\frac{1}{4k^2}\log\frac{\mu_+^2/4+k^2}{\mu_+^2/4+ \deta^2 k^2},
\end{align}
respectively, where $\mu_+^2\equiv 4\mu\mu_\text{ex}^2/m_\text{th}$ and $\deta\equiv |\Delta|/m_\text{th}$, with $\mu_\text{ex}^2\equiv \mex-\Delta^2$, $m_\text{th}\equiv m_1+m_2$, and $\Delta\equiv m_1-m_2$. 
Here $m_1$ and $m_2$ represent the masses of the scattering particles, and $m_{\rm ex}$ denotes the mass of the exchanged particle. The last line is obtained by taking the nonrelativistic approximation.
For $\eta\ll 1$ (e.g., for $DD^*$ scattering), $L_u(s)$ can be further expanded in powers of $\eta$, and the leading term takes the same form as $L_t(s)$ with $m_\text{ex}$ replaced by $\mu_+$. 

It is straightforward to deduce that the nearest lhc branch point is situated at $k_\text{lhc}^2 = -\mex/4$ and $-\mu_+^2/4$ for the $t$- and $u$-channel exchanges, respectively. For the $u$-channel exchange, as depicted in Fig~\ref{fig:feyndiag}~(b), we only consider the case $m_\text{ex}^2>\Delta^2$, otherwise the heavier of the two scattering particles can decay into the lighter one and the exchanged particle, leading to three-body dynamics with three on-shell particles beyond the scope of this Letter. 

Notably, the form of the lhc part of the $S$-wave in Eq.~\eqref{eq:L} remains unaffected by the coupling structures between the scattering and the exchanged particle. In the context of  $P$-wave couplings between the scattering and the exchanged particles, the lhc takes the form,
\begin{align}
\label{eq:pwd:P}
&\frac{1}{2}\int_{-1}^{+1}\frac{(\bold p_1-\bold p_3)^2}{t-\mex}\dd \cos \theta 
% = \frac{1}{2}\int_{-1}^{+1}\frac{-t}{t-\mex}\dd\cos\theta \\
= -\frac{\mex}{2}\int_{-1}^{+1}\frac{\dd\cos\theta}{t-\mex} - 1,\\
&\frac{1}{2}\int_{-1}^{+1}\frac{(\bold p_1-\bold p_4)^2}{u-\mex} \dd \cos \theta
\approx -\frac{\mu_\text{ex}^2}{2}\int_{-1}^{+1}\frac{\dd\cos\theta}{u-\mex} - 1, \nonumber
\end{align}
where the nonrelativistic approximation has been taken in the last line.
Vertices with higher partial waves can be treated similarly.

One way to integrate the lhc into the amplitude is to solve the Lippmann-Schwinger equation (LSE) with a potential that encompasses the long-range OPE potential responsible for the lhc, as demonstrated in, e.g., Refs.~\cite{Du:2023hlu,Meng:2023bmz}. The discontinuity of the $T$-matrix amplitude across the lhc associated with the OPE is the same as the OPE potential $V_\text{OPE}$, as only the tree-level diagrams in Fig.~\ref{fig:feyndiag} contribute to the lhc around the OPE lhc branch point. However, for a general potential, particularly one involving OPE, an analytical solution is infeasible and one has to resort to numerical solutions.

{\it The $N/D$ method.}---An alternative approach to incorporate both the lhc and the unitarity cut, i.e., the right-hand cut (rhc), is the $N/D$ method of dispersion relations~\cite{Chew:1960iv,Oller:2019rej}, in which the scattering amplitude can be expressed as a quotient of two functions, in the nonrelativistic kinematics,
\begin{equation}
f(k^2) = \frac{n(k^2)}{d(k^2)},
\end{equation}
where the denominator $d(k^2)$ only contains the rhc, and the numerator $n(k^2)$ has only the lhc but no rhc. It is immediately inferred from Eq.~\eqref{eq:invf} that these two functions satisfy
\begin{align}
\label{eq:im:nd}
\im d(k^2) & = - k\,n(k^2),   & \text{for } k^2&>0, \nno
\im n(k^2) & =  d(k^2)\,\im f(k^2),  & \text{for } k^2&< k^2_\text{lhc}.
\end{align}
Since $n(k^2)$ and $d(k^2)$ can be simultaneously multiplied by an arbitrary real analytic function without changing the amplitude, we set $n(k^2)$ free of poles, and thus the poles of $f(k^2)$ correspond to the zeros of $d(k^2)$. 

Making use of dispersion relations, one can write 
\begin{align} \label{eq:nd_disp}
n(k^2)&=n_m(k^2) + \frac{(k^2)^m}{\pi}\int_{-\infty}^{k^2_\text{lhc}}\frac{d(\kk)\im f(\kk)}{(\kk-k^2)(\kk)^m}\dd \kk,\\
d(k^2)&=d_n(k^2)-\frac{(k^2-k_0^2)^n}{\pi}\int_0^\infty \frac{k^\prime n(\kk)\, \dd\kk}{(\kk-k^2)(\kk-k_0^2)^n},\nonumber 
\end{align}
where $m$, $n$ are the numbers of subtractions required to render the integrals finite, and $n_m(k^2)$ and $d_n(k^2)$ are polynomials. The $N/D$ method has been demonstrated to be equivalent to solving the Lippmann-Schwinger equation, see e.g.~\cite{Oller:2018zts}. However, here we focus specifically on developing a practical parameterization of the amplitude with the lhc originating from the OPE, which could be very close to the threshold. Consequently, the imaginary part of the partial-wave amplitude, $\im f(k^2)$, is the same as that of the Feynman diagram in Fig.~\ref{fig:feyndiag}, i.e., 
\begin{equation}
    \im f(k^2) = c \im L(k^2) = -\frac{c }{4k^2}\pi, \quad \text{for } k^2<k_\text{lhc}^2,
\end{equation}
where $c$ is a parameter quantifying the strength of the lhc. It is proportional to the product of vertex couplings in Fig.~\ref{fig:feyndiag} and is contingent on the coupling structure of the vertices, as indicated in Eq.~\eqref{eq:pwd:P} for $P$-wave vertices. 

It is important to note that the integration over $d(\kk)$ in $n(k^2)$ is carried out along the lhc, where $d(\kk)$ is real analytic and can be expanded into a Taylor series in $k^2$. 
Then the dispersive integral in the $n(k^2)$ function in Eq.~\eqref{eq:nd_disp} is equivalent to $L(k^2)$ multiplied by a series plus a polynomial. 
One may divide both $n(k^2)$ and $d(k^2)$ by that series, modulo a normalization, to obtain (see End Matter for a detailed derivation)
\begin{equation}
    \label{eq:n}
n(k^2) = \tilde n(k^2) + \tg  (L(k^2)-L_0),
\end{equation}
where $\tilde n(k^2)$ is a rational function of $k^2$,  and the parameter $\tilde g$ is chosen such as the $\tilde n(k^2)$ is noramlized as $\tilde n(0)=1$. Here, $L_0=L(k^2=0)$ is subtracted to set the lhc term vanish at the threshold. It is straightforward to deduce from Eq.~\eqref{eq:L} $L_0 = -{1}/{\mex}$ and $-{1}/{\mu_\text{ex}^2}$ for the $t$- and $u$-channel exchanges, respectively.

The $d(k^2)$ function has the form 
\begin{equation}
d(k^2) = \tilde{d}(k^2)-ik (\tilde n(k^2)-\tg L_0) -\frac{\tg}{\pi}\int_0^\infty \frac{k^\prime L(\kk)}{\kk-k^2}\dd\kk, 
\end{equation}
where $\tilde{d}(k^2)$ is in general a rational function in $k^2$. It has a similar form of the first-iterated solution of the $N/D$ method in Refs.~\cite{Gulmez:2016scm,Oller:2019opk,Shen:2022zvd}, while in the latter case $n(k^2)$ is approximated from effective Lagrangians.

By substituting the $L(k^2)$ in Eq.~\eqref{eq:L} into the $d(k^2)$ function, one fortunately finds an analytical form,
\begin{align}
\label{eq:d}
d(k^2) = \tilde{d}(k^2)-ik\, n(k^2) -\tg d^\text{R}(k^2) ,
\end{align}
where $d^\text{R}(k^2)$ is given by 
\begin{align}
d^\text{R}_t(k^2) &= \frac{i}{4k}\log\frac{m_\text{ex}/2+ik}{m_\text{ex}/2-ik}, \nno
d^\text{R}_u(k^2) &= \frac{i}{4k}\left(\log\frac{\mu_+/2 +ik}{\mu_+/2 -ik}-\log\frac{\mu_+/2 +i\deta k }{\mu_+/2 - i\deta k } \right),
\end{align}
for the $t$- and $u$-channel exchanges, respectively.  
It is worth stressing that $d(k^2)$ is free of lhc, as the lhc associated with $n(k^2)$ below the threshold is counterbalanced by $d^R(k^2)$, which is crucial to ensure that $f(k^2)$ exhibits the correct lhc behavior~\footnote{We have explicitly checked that the lhc behavior is severely violated and one cannot describe the phase shift from the $T$-matrix without $d^\text{R}$.}. Along the rhc, both $n(k^2)$ and $d^\text{R}(k^2)$ are real such that $\im d(k^2)=-k\, n(k^2)$. 

{\it ERE with the left-hand cut.}---Using Eqs.~\eqref{eq:n} and \eqref{eq:d}, the scattering amplitude can then be parameterized as
\begin{equation}
\label{eq:f}
\frac{1}{f_{[m,n]}(k^2)} = \frac{\sum_{i=0}^n\tilde{d}_ik^{2i}- \tg d^\text{R}(k^2)}{1 +\sum_{j=1}^m\tilde{n}_j k^{2j} + \tg(L(k^2)-L_0)}-ik,
\end{equation}
where $\tilde g$, $\tilde d_i (i=0,\ldots,n)$ and $\tilde n_j (j=1,\ldots,m)$ are parameters, and the rational functions $\tilde n(k^2)$ and $\tilde{d}(k^2)$ have been expanded into polynomials. This form will be called an $[m,n]$ approximant and can be regarded as an extension of ERE with the lhc. 
Note that the nonrelativistic approximation discussed above uses $E = k^2/(2\mu)$, and the approximant can be at most $[1,1]$ for self-consistency. However, the method can be straightforwardly generalized by keeping higher terms in nonrelativistic expansion, which then allows for higher approximants.

In the absence of OPE, i.e., $\tg=0$, the $[1,1]$ approximant reduces to the conventional ERE in Eq.~\eqref{eq:ere0}, as the $\tilde{n}_1$ term can be absorbed by redefining $\tilde{d}_1$ up to higher orders. 

The parameters in Eq.~\eqref{eq:f} need to be determined by experimental data or lattice QCD simulations. 
In the following, we focus on the $[0,1]$ approximant, 
\begin{equation}
    \label{eq:final}
{f_{[0,1]}(k^2)} = \left[\dfrac{\tilde d_0+ \tilde d_1 k^2- \tg d^\text{R}(k^2)}{1 + \tg(L(k^2)-L_0)}-ik\right]^{-1}.
\end{equation}

One can deduce the $S$-wave scattering length $a=f(k^2=0)$ and effective range $r=\frac{\dd^2(1/f+ik)}{\dd k^2}\Big|_{k=0}$ from Eq.~\eqref{eq:final}. 
It is easy to find that~\footnote{An extra term $-2\tilde{n}_1/a$ is present for the [1,1] approximant, where $a$ denotes $a_t$ and $a_u$ for the $t$- and $u$-channel exchanges, respectively. } for the $t$-channel exchange,

\begin{align}
\label{eq:ar:t}
a_t  = \left(\tilde{d}_0+\frac{\tg}{m_\text{ex}}\right)^{-1}, \quad
r_t = 2\tilde{d}_1 -\frac{8\tg}{3m_\text{ex}^3} -\frac{4\tg}{m_\text{ex}^4 a_t} ,
\end{align}
and for the $u$-channel exchange,
\begin{align}\label{eq:ar:u}
a_u &=\left[\tilde d_0+\frac{\tg}{\mu_+}\left(1- \deta\right)\right]^{-1}, \nonumber\\
r_u &= 2\tilde{d}_1 - \frac{8\tg}{3\mu_+^3} \left(1- \deta^3\right) -\frac{4\tg}{ \mu_+^4 a_u}\left(1- \deta^4\right). 
\end{align}

\begin{figure}[tb]
    \begin{center}
    \includegraphics[width=\linewidth]{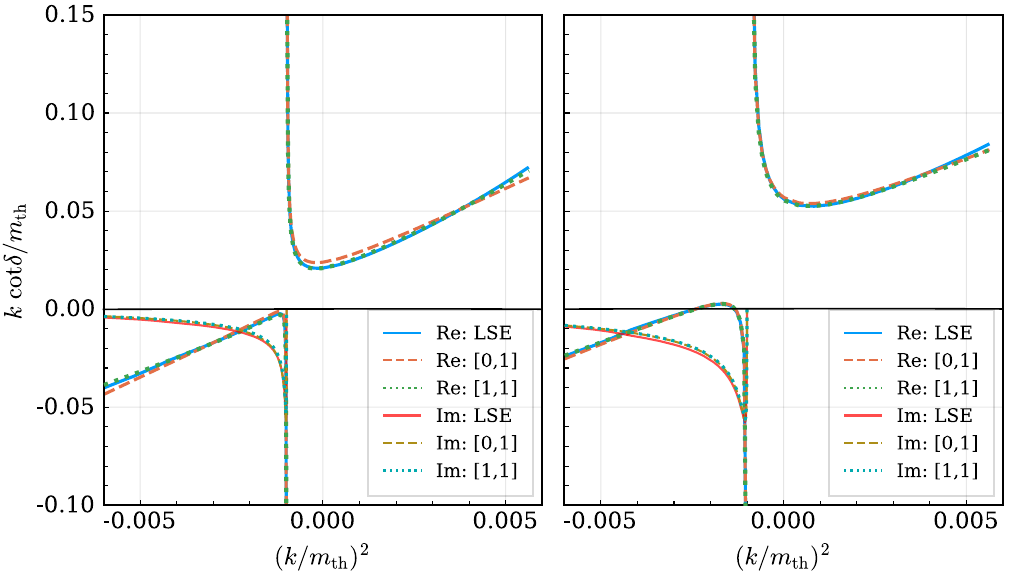}
    \caption{Comparison between results from the $[0,1]$ approximant in Eq.~\eqref{eq:final} (denoted as ``$[0,1]$"), the $[1,1]$ approximant (denoted as ``$[1,1]$") and the LSE solutions in Ref.~\cite{Du:2023hlu}. 
    The two panels correspond to the two fits in Fig.~3 in Ref.~\cite{Du:2023hlu}. 
    The solid curves represent the real (blue) and imaginary (red) parts of $k\cot\delta$ solved from the LSE, and the dash (dotted) curves correspond to the real and imaginary parts from Eq.~\eqref{eq:final} (the $[1,1]$ approximant), with the latter restricted to the $k^2<0$ region. Although only the real parts of the phase shift are employed in the fitting, the imaginary parts along the lhc obtained in our parameterization are almost indistinguishable from those of the LSE. 
    }
    \label{fig:phaseshift}
    \end{center}
\end{figure}

\begin{figure}[tb]
    \begin{center}
    \includegraphics[width=0.48\textwidth]{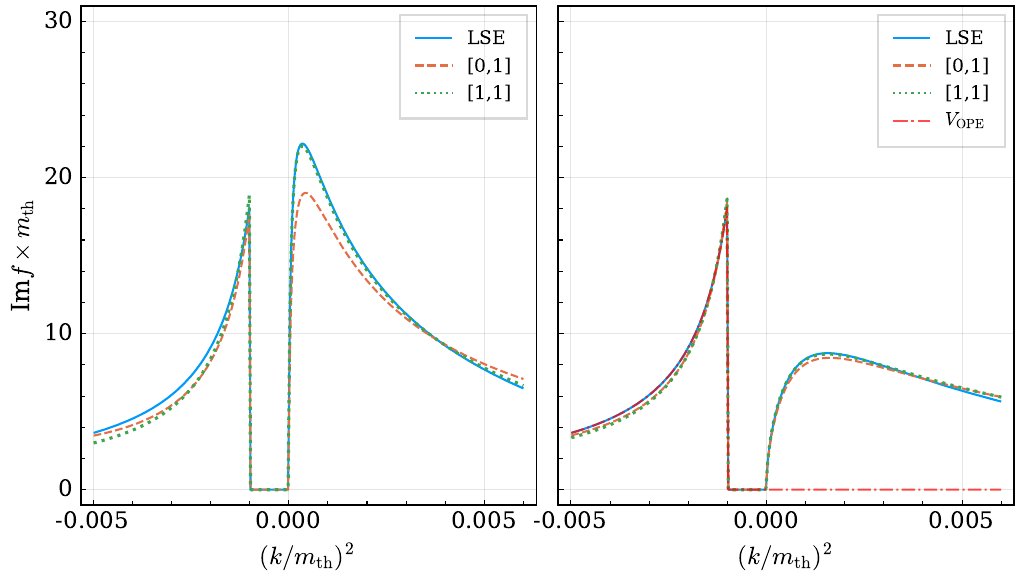}
    \caption{Comparisons of $\im f$ from  %Eq.~\eqref{eq:final} 
    the $[0,1]$ and $[1,1]$ approximants
    and that from the LSE. The two panels correspond to those in Fig.~\ref{fig:phaseshift}. $\im f$ along the lhc from LSE are the same in both panels and is equal to $-\mu/(2\pi)\im V_\text{OPE}$, shown in the right panel for comparison. 
    }
    \label{fig:imf}
    \end{center}
\end{figure}

{\it Example.}---As an illustrative example, we attempt to employ Eq.~\eqref{eq:final} to reproduce the amplitudes of the $DD^*$ system solved using LSE in Ref.~\cite{Du:2023hlu}, for which the lhc from the $u$-channel one-pion exchange appears near the $DD^*$ threshold. 

We take 6 equidistant pseudo-data points generated by the real part of the phase shift $k\cot\delta$ in Fig.~3 of \cite{Du:2023hlu} for $(k/m_\text{th})^2\in [-0.005,0.005]$ with uniform uncertainties, and fit them using Eq.~\eqref{eq:final} with three parameters ($\tilde d_0$, $\tilde d_1$ and $\tilde g$). 
The results are shown in Fig.~\ref{fig:phaseshift}. 
One sees that both the real and imaginary parts of the phase shift from Eq.~\eqref{eq:final} match the LSE solutions remarkably well.
In particular, the lhc and the zero of the amplitude (divergence of $k\cot\delta$) are well captured by Eq.~\eqref{eq:final}.

Finally, we compare the imaginary part of scattering amplitude from LSE with Eq.~\eqref{eq:final} in Fig.~\ref{fig:imf}, from which it is evident that Eq.~\eqref{eq:final} captures the correct behavior of the lhc. 
The agreement can be improved by using higher approximants as can be seen from the $[1,1]$ curves in Figs.~\ref{fig:phaseshift} and \ref{fig:imf}.

{\it Couplings to the exchanged particle.}---As demonstrated in Fig.~\ref{fig:imf}, since the discontinuity of the amplitude arising from the OPE is just the tree-level OPE potential $V_\text{OPE}$ iterated in the LSE, it is easy to infer $\im f = -\frac{\mu}{2\pi}\im T=-\frac{\mu }{2\pi}\im V_\text{OPE}$ along the lhc. Therefore, if sufficient experimental data or lattice QCD results are available to fix the parameters in Eq.~\eqref{eq:final}, we can extract the product of the couplings in Fig.~\ref{fig:feyndiag}. 

Along the lhc, we have 
\begin{equation}
    \im V_{\rm OPE}(k^2) = g_P\frac{-\pi}{4k^2}\mathcal{F}_\ell, \quad \text{for } k^2<k_\text{lhc}^2,
\end{equation}
where $g_P\equiv g_1g_2$ ($g^2$) for the $t$-channel ($u$-channel) exchange, 
and $\mathcal{F}_\ell$ is a factor related to the coupling structures. If both vertices are in $S$ waves, then $\mathcal{F}_\ell = 1$. If both are in $P$ waves, $\mathcal{F}_\ell = -{\mex}$ or $-{\mu_\text{ex}^2}$, depending on the $t$- or $u$-channel exchange, c.f., Eq.~\eqref{eq:pwd:P}. Similarly, the $n(k^2)$ function obeys
\begin{equation}
   \im n(k^2) = -\tg \frac{\pi}{4k^2}, \quad \text{for } k^2<k_\text{lhc}^2. 
\end{equation}
The $d(k^2)$ function is real and analytic along the lhc, we thus evaluate it at the branch point of the lhc~\footnote{An extra term $-\tilde{n}_1 m_\text{ex}^3/8$ ($-\tilde{n}_1 \mu_\text{+}^3/8$) is to be added for the $t$-exchange ($u$-exchange) in the [1,1] approximants.},
\begin{align}
    d_t^{0,{\rm lhc}} &=\tilde d_0 - \frac{\tilde{d}_1\mex}{4}+\frac{m_\text{ex}}{2}\left(1+\frac{\tg}{\mex}\right)+\frac{\tg \log 2 }{m_\text{ex}} , \nonumber\\
    d_u^{0,{\rm lhc}} &= \tilde d_0-\frac{\tilde d_1 \mu_+^2}{4}+\frac{\mu_+}{2}\left(1+\frac{\tg}{\mu_\text{ex}^2}\right) +\frac{\tg \log[2/{(1+\deta)}]}{\mu_+}. 
    \label{eq:dlhc}
    \end{align}
Then we equate the imaginary parts of the $f$ and $-\frac{\mu}{2\pi}V_\text{OPE}$ at the branch point of the lhc,
\begin{equation}\label{eq:gg}
    g_P = -\frac{2\pi\tg}{\mu d^{0,\text{lhc}}\mathcal{F}_\ell}.
\end{equation}
With the parameters fixed in the fit shown in Fig.~\ref{fig:phaseshift} and ultilizing Eq.~\eqref{eq:gg}, we find ${g_{D^*D\pi}^2}/{(4F^2)}=9.0~\text{GeV}^{-2}$ and $9.4~\text{GeV}^{-2}$, respectively, fully in consistent with the value $9.2~\text{GeV}^{-2}$ used in Ref.~\cite{Du:2023hlu}, where $g_{D^*D\pi}$ is the $D^*D\pi$ coupling constant and $F$ is the pion decay constant. 

Conversely, if the couplings of the exchanged particle to the scattering particles are known in advance, one can reduce the number of free parameters in the parameterization via Eq.~\eqref{eq:gg}. 
It is particularly useful when the available data points are limited.

{\it Amplitude zeros.}---Amplitude zeros~\cite{Castillejo:1955ed} (or poles of $k\cot\delta$) have been found in scatterings involving three-body dynamics, such as in the $DD^*$ scattering~\cite{Du:2023hlu} (see Fig.~\ref{fig:phaseshift}) and in the scattering of neutron off one-neutron halo nuclei~\cite{Yamashita:2008sg, Shalchi:2016psb, Deltuva:2017zvk,Zhang:2023wdz}.
Such zeros are suggested to be a signal of an excited Efimov state (for a review of Efimov physics, see Ref.~\cite{Braaten:2004rn}) close to the threshold~\cite{Yamashita:2008sg, Zhang:2023wdz}.
These scatterings correspond to the $u$-channel exchange processes discussed here.
From our parameterization, it is clear that both the $t$-channel and $u$-channel exchanges can lead to amplitude zeros, as solutions of $n(k^2)=0$.

At leading order, {i.e., $\tilde{n}(k^2)=1$, the amplitude has a zero for $y>0$ for the $t$-channel exchange, and the zero is located at 
\begin{align}
k^2_{t,\text{zero} }= -\frac{\mex}{4} \left[1+ \frac{1}{y} W(-e^{-y}y)\right],
\label{eq:kzero}
\end{align}
where $y\equiv  1+\mex/\tg$ and $W$ is the Lambert $W$ function. The Lambert $W$ function has two branches for real $y$. In the case of $\tg>0$, i.e., $y>1$, the principal branch $W_0$ is selected. For $\tg <-\mex$, where $0<y<1$, the other branch $W_{-1}$ is chosen.}
The location of the zero is due to the interplay of the long-range (from the OPE) and short-range (from the $\tilde d_{0,1}$ terms in $\tg$, c.f., Eq.~\eqref{eq:gg}) interactions.
Using Eqs.~\eqref{eq:ar:t} (\eqref{eq:ar:u} for the $u$-channel exchange) and \eqref{eq:gg}, one may express $\tg$ in terms of couplings, scattering length and effective range. Then, we have
\bea\label{eq:y}
y= 1 + \frac{1+\frac{4}{3}a_t m_\text{ex}(1-\log 4)-\frac{4\pi a_t \mex}{\mu g_P \mathcal{F}_\ell}   }{2+a_t m_\text{ex}\left(1-m_\text{ex}r_t/4\right)}. 
\eea

For a general $u$-channel exchange, the zero is determined by the solution of 
\begin{align}
    1+\tilde{g}\left[L_u\left(k^2_{u,\text{zero}}\right)+\frac{1}{\mu_{\mathrm{ex}}^2}\right] = 0,
    \label{eq:uzero}
\end{align}
which does not permit a simple analytic expression.
However, for the case $|\Delta| \ll m_\text{th}$ such that $\deta\ll 1$, all the expressions derived above for the $t$-channel exchange apply to the $u$-channel case with $m_\text{ex}$ replaced by $\mu_+$. 
We have checked that the zero obtained in this way indeed agrees with the divergent point of $k\cot\delta$ shown in Fig.~\ref{fig:phaseshift}.

{\it Summary.}---In summary, we have proposed a new model-independent parameterization of the scattering amplitude at low energies in the form of Eq.~\eqref{eq:f}. 
It is like the Pad\'e approximation but with explicitly nonanalytic terms.
The parameterization incorporates the lhc from the long-range interaction due to the $t$- and $u$-channel particle exchanges, which strongly restricts the validity range of the traditional ERE. 
Our parameterization reduces to the traditional ERE in the absence of lhc. 

We have shown that our parameterization in the simple $[0,1]$ approximant form indeed reproduces the $DD^*$ low-energy scattering amplitudes from the LSE in Ref.~\cite{Du:2023hlu} remarkably well; especially, it captures the correct behavior due to the lhc.
The $[0,1]$ approximant has only three parameters; one of them can be fixed in advance if the couplings of the exchanged particle to the scattering particles are known, and then the number of parameters is the same as that of the traditional ERE up to $\mathcal{O}(k^2)$.

The new parameterization can be used for analyzing a wide class of low-energy scatterings that involve a long-range (yet finite-range) interaction across different fields of physics. For instance, neutron-deuteron, $^3$He-$\alpha$ scatterings and scatterings involving a halo nucleus~\cite{Hammer:2022lhx,Zhang:2023wdz} in nuclear physics, dimer-atom (e.g., $^4$He atom-$^4$He$_2$ dimer~\cite{Kolganova:2011uc}) interactions in cold atom physics~\cite{Braaten:2004rn}, and the self-interacting dark matter with Yukawa-type interactions via a light mediator proposed to explain the apparent mass deficit in astrophysical small-scale halos~\cite{Chu:2019awd,Kamada:2023iol}. 
In particular, the scattering between an atom and a dimer composed of two atoms with an exceptionally small binding energy can generate a dominant long-range interaction mediated by the $u$-channel (nearly on-shell) atom-exchange diagram (dimer recombination), which leads to an lhc in the vicinity of the atom-dimer threshold. 
The proposed parameterization provides a practical method to study such systems, which would otherwise require to solve the Schr\"odinger or Faddeev equations.
The long-range interaction is also prevalent in many hadron-hadron interactions, such as baryon-(anti)baryon, $B^{(*)}B^*$, $B^{(*)}\bar{B}^*$, $ND^{*}$, $N\bar D^{*}$, $\Sigma_{(c)}^{(*)}D^{*}$, $\Sigma_{(c)}^{(*)}\bar D^{*}$, and so on, at physical or unphysical (relevant for lattice QCD) quark masses. In particular, it can be applied to the nucleon-nucleon, $DD^*$ and $D\bar{D}^*$, that have been extensively investigated in lattice QCD calculations (Refs.~\cite{Green:2021qol,Padmanath:2022cvl} as recent examples), where the traditional ERE without the lhc has been shown to be problematic~\cite{Du:2023hlu}. The explicit inclusion of the lhc can be crucial in accurately determining pole positions. The new parameterization is expected to be invaluable for understanding various near-threshold hadron resonances. 

We also derive amplitude zeros, which are determined by the interplay of the long- and short-range interactions, in terms of low-energy observables. 
Without this parameterization, such zeros can only be properly treated by solving the much more complicated integral equations. In certain scenarios, more than one OPE diagram may contribute to the lhc, e.g. both $t$- and $u$-channel OPE are present. The corresponding $n(k^2)$ and $d(k^2)$ can be modified by adding a term $\tilde{g}(L(k^2)-L_0)$ to $n(k^2)$, and a term $\tilde{g}d^R(k^2)$ to $d(k^2)$, respectively, for each OPE diagram.

\begin{acknowledgments}
FKG acknowledges valuable discussions with Hans-Werner Hammer and Tao Shi regarding potential applications of this framework to cold atom systems. This work is supported in part by the National Key R\&D Program of China under Grant No.~2023YFA1606703; by the Chinese Academy of Sciences under Grant No.~YSBR-101; and by the National Natural Science Foundation of China under Grants No. 12125507, No. 12361141819, and No. 12447101.

\end{acknowledgments}

\bibliography{refs}

\begin{onecolumngrid}

\section*{End Matter}

\end{onecolumngrid}

\begin{twocolumngrid}

In this End Matter, we give a step-by-step derivation of Eq.~\eqref{eq:n} in the main text. By construction, the $d(k^2)$ is analytic along the lhc and thus can be parameterized as a polynomial,
\begin{align}
P(k^2) & =  \sum_{i=0}  d^\prime_i(k^2+k_0^2)^i \notag\\
&= \sum_{i=0} d_i(k^2)^i = d_0 + d_1 k^2 + d_2(k^2)^2+\dots ,
\end{align}
where $(-k_0^2)$ is an arbitrary point along the lhc around which $d(k^2)$ is expanded, $d_i^{\prime}$ are coefficients, and $d_i$ are the coefficients of the rearranged polynomial. Then one has 
\bea
n(k^2) &=& n_m(k^2) + \frac{(k^2)^m}{\pi}\int_{-\infty}^{k^2_\text{lhc}}\frac{d(\kk)\im f(\kk)}{(\kk-k^2)(\kk)^m}\dd \kk \nonumber\\
&=& \sum_{i=0}\left[  \bar n_i(k^2)^i + \frac{(k^2)^i}{\pi} \int_{-\infty}^{k^2_\text{lhc}} \frac{d_i (\kk)^i\im f(\kk) }{(\kk -k^2)(\kk)^i}\dd \kk\right] \nonumber\\ 
&=& \left(\sum_i\bar{n}_i(k^2)^i\right) + \frac{\sum_i d_i(k^2)^i}{\pi }\int_{-\infty}^{k^2_\text{lhc}}\frac{\im f(\kk)}{\kk-k^2}\dd \kk \nonumber\\
&=& \bar{n}(k^2) + cP(k^2) (L(k^2) -L_0),
\eea
where $\bar{n}(k^2)=\sum_i\bar{n}_i(k^2)^i + c L_0 P(k^2)$ is a polynomial, with $\bar{n}_i$ coefficients of polynomials. Redefining $n(k^2)$ by dividing the polynomial $cP(k^2)/\tg$, which can always be done since the same polynomial can be divided simultaneously from $d(k^2)$, one obtains Eq.~\eqref{eq:n}
\bea 
n(k^2) = \tilde{n}(k^2)+\tg(L(k^2)-L_0),
\eea 
where $\tilde{n}(k^2) = \tg \bar{n}(k^2)/(cP(k^2))$ is a rational function.

The crucial observation for the parameterization is that the imaginary part of the amplitude along the OPE lhc stems from the OPE tree-level diagram, i.e., $\text{Im}T = \text{Im}V_\text{OPE}$. It is easy to check from the Lippmann-Schwinger equation that the difference between the on-shell partial wave amplitude and the tree-level OPE potential,
\begin{align}
    T(E,p,p)-V(E,p,p)=\int \frac{\text{d}^3\vec{k}}{(2\pi)^3}\frac{V(E,p,k)T(E,k,p)}{E-k^2/2\mu}
\end{align}
is free of OPE lhc on the first Riemann sheet, where $p=\sqrt{2\mu E}$ is the on-shell momentum. 
The reason is that the branch point of the lhc is below threshold, while the integration contour is above threshold and thus the integral is free of singularities on the first Riemann sheet (the lhc will appear on unphysical Riemann sheets, which, however, does not matter for the ERE).
Thus, the iterated OPE ladder diagrams do not contribute to the lhc relevant for generalizing the ERE.
%  as the corresponding singularities of on-shell OPE are located on unphysical Riemann sheets. 

\begin{figure}[tb]
\begin{center}
\includegraphics[width=\linewidth]{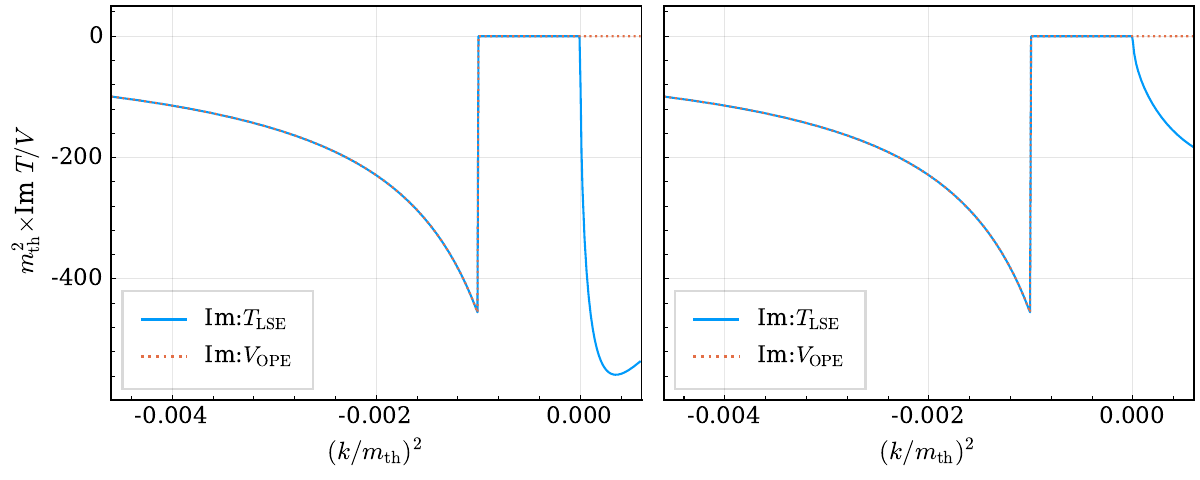}
\caption{The imaginary part of the potential (dotted curve) and that of the amplitude from the Lippman-Schwinger equation (solid curve). The left and right panels correspond to the left and right ones in Fig.~\ref{fig:phaseshift} or \ref{fig:imf}, respectively.  }
\label{fig:V_vs_T}
\end{center}
\end{figure}

In order to check the above statement explicitly, we plot the imaginary part of the tree-level OPE potential $V_\text{OPE}$ and that of the amplitude from the Lippmann-Schwinger equation in Fig.~\ref{fig:V_vs_T}. One can clearly see that the on-shell scattering amplitude $T$ has exactly the same imaginary part as the tree-level OPE potential $V_\text{OPE}$ along the lhc.

\end{twocolumngrid}

\end{document}